\documentclass[aps,twocolumn,nofootinbib,preprintnumbers,prl,10pt]{revtex4-2}
\usepackage{amsfonts,amsthm,amsmath,amssymb,graphicx,hyperref}
\usepackage{braket,url,color}

\newcommand{\linefill}{
	{-}\mkern-7mu
	\cleaders\hbox{$\mkern-2mu-\mkern-2mu$}\hfill
	\mkern-7mu{-}%
}
\usepackage{subfigure}

\begin{document}

\title{Eigenstate capacity and Page curve in fermionic Gaussian states}

\author{Budhaditya Bhattacharjee$^{1}$\email{(In alphabetical order) budhadityab@iisc.ac.in}, Pratik Nandy$^{2}$\email{pratiknandy@iisc.ac.in} and Tanay Pathak$^{3}$\email{tanaypathak@iisc.ac.in}}
\affiliation{
 Centre for High Energy Physics, Indian Institute of Science, C.V. Raman Avenue, Bangalore, India. }

\begin{abstract}
Capacity of entanglement (CoE), an information-theoretic measure of entanglement, defined as the variance of modular Hamiltonian, is known to capture the deviation from the maximal entanglement. We derive an exact expression for the average eigenstate CoE in fermionic Gaussian states as a finite series, valid for arbitrary bi-partition of the total system.  Further, we consider the complex SYK$_2$ model in the thermodynamic limit and we obtain a closed-form expression of average CoE. In this limit, the variance of the average CoE becomes independent of the system size. Moreover, when the subsystem size is half of the total system, the leading volume-law coefficient approaches a value of $\pi^2/8 - 1$.  We identify this as a distinguishing feature between integrable and quantum-chaotic systems.  We confirm our analytical results by numerical computations. 
\end{abstract}

\maketitle

\section{Introduction}
Ranging from critical phenomena \cite{PhysRevLett.90.227902} to the black hole information problem \cite{Hawking:1975vcx}, the notion of entanglement is ubiquitous. For example, in the latter case, one typically considers the time evolution of the entanglement entropy (EE) between the outgoing Hawking quanta and the interior of the black hole, which follows the Page curve \cite{RevModPhys.93.035002}. It is known that the implications and appearance of the Page curve are far more general; it holds for bi-partition of any quantum system. In particular, Page showed \cite{PhysRevLett.71.3743} (see \cite{PhysRevE.52.5653, PhysRevLett.72.1148, PhysRevLett.77.1} for the proof) that, for a random pure state in the thermodynamic limit, a subsystem has nearly maximal entropy. 

Recently, similar considerations are explored for the fermionic Gaussian states \cite{PhysRevB.103.L241118}, where an exact expression for the average EE is obtained. In the thermodynamic limit, the average EE appears to be a function of the subsystem fraction. It contrasts with general quantum-chaotic Hamiltonians, where to the leading order, the average EE is independent of the subsystem fraction \cite{PhysRevLett.119.220603}. As a consequence, along with many other recent observations \cite{PhysRevE.81.036206, Deutsch_2010, PhysRevE.89.012125, Beugeling_2015, PhysRevX.8.021026, PhysRevLett.119.020601, HUANG2019594, PhysRevE.99.032111, Nakagawa:2017yiw, PhysRevD.100.041901, PhysRevB.97.245126, PhysRevLett.121.220602, PhysRevB.99.075123, Fujita:2018wtr, PhysRevD.100.105010, PhysRevLett.127.040603,  PhysRevE.100.022131, PhysRevE.100.062134, PhysRevLett.125.180604, HUANG2021115373, PhysRevB.103.104206,  PhysRevResearch.3.023213, DePalma:2021cbf, PhysRevA.103.062416, kumari2021eigenstate, PhysRevE.104.014146}, it has been argued that the leading volume-law coefficient of the average EE can distinguish integrable and chaotic systems \cite{Caux_2011}. One question naturally arises - is there any other entanglement measure whose properties are different from EE and R\'enyi entropies (RE) but can still differentiate between integrable and quantum-chaotic systems?

In this paper, we address this question and find an affirmative answer. We consider a different entanglement measure, the capacity of entanglement (CoE) \cite{Nakaguchi:2016zqi, PhysRevD.99.066012, deBoer:2020snb}, first considered in the Kitaev model to describe topologically ordered states \cite{PhysRevLett.105.080501} and to understand the thermodynamic properties of the entanglement spectrum \cite{PhysRevB.96.205108}. Very recently, it has received considerable attention, for example, in local operator excitations \cite{Nandy:2021hmk}, random pure states \cite{Okuyama:2021ylc}, and in the context of holography and black hole information problem, where it is shown to capture partially entangled states in replica wormhole geometry \cite{Kawabata:2021hac, Kawabata:2021vyo}. While the EE is the expectation value of the modular Hamiltonian $K_A = -\log \rho_A$, the CoE is defined as the variance of $K_A$, namely \cite{PhysRevD.99.066012, Nandy:2021hmk}
\begin{align}
C_A = \braket{K_A^2} - \braket{K_A}^2 = \lim_{n \rightarrow 1} n^2 \partial_n^2 \ln \mathrm{Tr} \rho_A^n. \label{capdef}
\end{align}
Here $\rho_A$ is the reduced density matrix of the subsystem $A$, and $n$ is the R\'enyi index. The second equality comes from the replica method and is useful for holographic computations \cite{Nakaguchi:2016zqi}. It has been shown that in the case of operator excitations, the departure from the maximally entangled state can be captured by its CoE, which can be interpreted as entanglement between quasi-particles \cite{Nandy:2021hmk}.

This motivates us to consider the CoE in fermionic Gaussian states, with a broader motivation to see how far it can be used as a tool to characterize integrable and chaotic systems. In this paper, we derive an exact expression for the average eigenstate CoE for arbitrary bi-partition of the full system as a finite series. We show that the average CoE follows the volume-law, and the thermodynamic limit is approached from below, contrary to the behaviour of the average EE found in \cite{PhysRevB.103.L241118}. The behaviour of the average CoE (density) also differs from that of the average EE (density) in small subsystem size. It is convex with respect to the subsystem fraction $f$ and vanishes in the limit $f \rightarrow 0$. The justification is that as one approaches the limit of the vanishingly small subsystem, the average CoE also vanishes due to maximal entanglement. 

In this context, we study the complex $\mathrm{SYK}_2$ model both analytically and numerically in the thermodynamic limit, and we derive an analytic expression for the average CoE in a closed-form expression. Moreover, in this limit, the variance of the average CoE becomes independent of the system size. When the subsystem size is half of the full system, the leading volume-law coefficient approaches a value of $\pi^2/8 - 1$. We recognize this as a distinguishing feature between integrable and quantum-chaotic systems.

\section{Capacity for fermionic Gaussian states}

In this section, we calculate the average eigenstate CoE in fermionic Gaussian states \footnote{Fermionic Gaussian states also play an important role in various other information-theoretic quantities like circuit complexity. See \cite{PhysRevD.98.126001, Hackl:2018ptj} and the references therein.}. Since Gaussian states lie in the sub-manifold of pure states and are related by Bogoliubov transformation $W \in O(2N)$ \cite{PhysRevB.103.L241118}, the average CoE is defined as
\begin{align}
\braket{C_A}_G = \int_{W \in O(2N)} \mathrm{d W} ~C_A (\ket{J_W}),
\end{align}
where the integration is over all Gaussian states $\ket{J_W}$. Following \cite{PhysRevB.103.L241118}, one can use the complex structure of $J$, and compute $C_A (\ket{J_W}) = \sum_{j=1}^{V_A} c(x_i)$, where $V_A$ is the subsystem size. From \cite{Nandy:2021hmk}, we obtain the expression for $c(x_i)$ as
\begin{align}
c(x_i) = \frac{1}{4}(1-x_i^2) \bigg[\ln \bigg(\frac{1+x_i}{1-x_i}\bigg)\bigg]^2. \label{cmain}
\end{align}
The average CoE is calculated as $\braket{C_A}_G = V_A \int_0^1 \mathrm{dx} \, c(x) \rho(|x|) = \tfrac{V_A}{2} \lim_{\epsilon \rightarrow 0} \, \partial^{2}_{\epsilon} I_{\epsilon}$,
where we have considered the integral with parameter $\epsilon$
\begin{equation}
I_{\epsilon}=  \frac{1}{8}\int_{-1}^{1}\mathrm{d x} \, (1-x^2)  \, \bigg(\frac{1+x}{1-x}\bigg)^{\epsilon}  \rho(|x|), \label{cc1}
\end{equation}
and the density function is given by the distribution \cite{PhysRevB.103.L241118}.
\begin{align}\label{dis}
\rho(|x|) = \frac{(1-x^2)^{\Delta}}{V_A}\sum_{k =0}^{V_A -1} \frac{[\mathcal{P}^{(\Delta, \Delta)}_{2 k} (|x|)]^2}{c_k}.
\end{align}
Here $\mathcal{P}^{(\Delta, \Delta)}_{2 k} (|x|)$ is the Jacobi polynomial and  $\Delta = V - 2 V_A \geq 0$, where $V$ is the size of the full system. The coefficients $c_k$ are given by  
$c_k = 2^{2 \Delta} [(2k + \Delta)!]^2/[(2 k)! (2 k + 2 \Delta)! (4 k + 2 \Delta +1)]$ \cite{PhysRevB.103.L241118}.
The integral \eqref{cc1} can be evaluated as a finite series, by using a series representation of Jacobi polynomials. After some simplifications (see Appendix A), we arrive at the following expression for the average CoE
\begin{align}
\braket{C_A}_{G} &=\sum_{j=0}^{V_{A}-1} \frac{(2 \Delta +4 j+1) \Gamma (\Delta +2)}{\Gamma (2 j+1) \Gamma (2 j+2 \Delta +1)} \nonumber \\ &\times \sum_{m,k=0}^{2j} \mathcal{G}_1 (k) \, \mathcal{G}_1 (m) \, \mathcal{G}_2 (k + m), \label{pandora}
\end{align}
where the functions $\mathcal{G}_1(p)$ and $\mathcal{G}_2 (q)$ are given by
\begin{align}
\mathcal{G}_1 (p) &= (-1)^p \,\binom{2 j}{p}  \frac{\,\Gamma (2 j+p+2 \Delta +1) }{\Gamma (p+\Delta +1)}, \nonumber  \\
\mathcal{G}_2 (q) &= \frac{\Gamma(q + \Delta + 2)}{\Gamma (q + 2 \Delta + 4)}\nonumber \Big( \big[ \Psi(q + \Delta + 2) - \Psi(\Delta +2)\big]^2  \nonumber \\ &\;\;\;\;\;\;\;\;\;\;\;\;\; \;\;\;\;\;\;\; + \Psi_1 (q + \Delta + 2) + \Psi_1 (\Delta + 2) \Big).
\end{align}
Here $\Psi(z)$ and $\Psi_1(z)$ are the digamma and trigamma functions respectively. Note that, the expression \eqref{pandora} is a general result which holds for arbitrary bi-partition of the full system. It is difficult to vizualize what happens at the thermodynamic limit. However, we can directly consider the thermodynamic limit of the ensemble \eqref{dis}, and obtain the expression of CoE in this limit. We will derive this result in later part of the paper.

\begin{figure}
	\centering
	\includegraphics[height=5cm,width=1\linewidth]{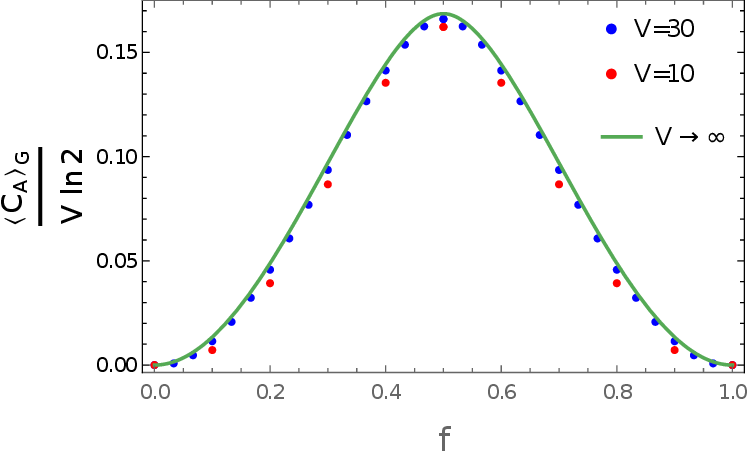}
	\caption{Page curve of the average CoE in fermionic Gaussian states. The plots are obtained using the expression \eqref{pandora} for $V=30$ (blue) and $V=10$ (red) respectively. The continuous curve (green) is in the thermodynamic limit.}
	\label{fig:page}
\end{figure}

For finite $V$ and $V_A$, the behaviour of the average CoE with respect to subsystem size is shown in Fig.\ref{fig:page}. We refer it as the ``Page curve" for the average CoE. The dotted points are obtained using \eqref{pandora}, while the continuous curve is the exact expression in the thermodynamic limit, given by \eqref{analy}. For small $f= V_A/V$, the average CoE is a convex function of $f$, as seen from Fig.\ref{fig:page}.  We also note that the thermodynamic limit is obtained from below. These two properties are in sharp contrast to that of the average EE as found in \cite{PhysRevB.103.L241118}.

In Fig.\ref{fig:dev} (a), we have shown the deviation of the average CoE for a finite system size from its thermodynamic limit. The intercepts of linear-fit satisfy $|a_1| \leq 10^{-4}$, which implies that the deviation becomes small as we approach the thermodynamic limit.

\section{Calculation of variance}

\begin{figure}
	\subfigure[]{\includegraphics[height=3cm,width=0.47\linewidth]{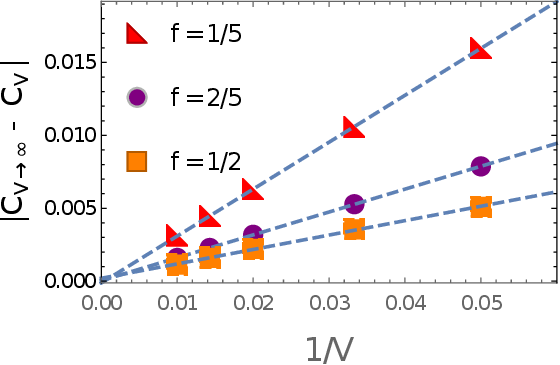}}
	\hfill
	\subfigure[]{\includegraphics[height=3cm,width=0.47\linewidth]{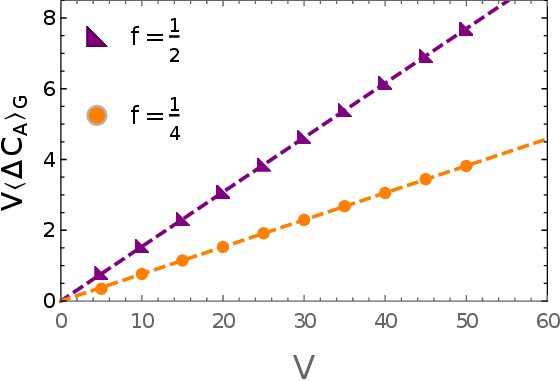}}
	\caption{(a) Deviation of finite $V$ result from the result at the thermodynamic limit (Eq.\eqref{analy}). The plots are done for finite subsystem fraction, $f=1/5 ~(\mathrm{red}), 2/5 ~(\mathrm{purple})$ and $1/2 ~(\mathrm{orange})$ respectively. For all cases, the intercepts ($a_1$) satisfy $|a_1| \leq 10^{-4}$. (b) The standard deviation of the average CoE times the volume with respect to the full volume.} \label{fig:dev}
\end{figure}

To see whether the average CoE is also typical, it is instructive to calculate the variance. The variance is defined as
\begin{align*}
(\Delta C_A)_G^2 = \int_0^1 \mathrm{dx} \,c^2(x) \mathcal{K}_{x,x} - \int_0^1 \mathrm{d^2 x} \,c(x_1) c(x_2) \mathcal{K}_{x_1,x_2}, \label{var}
\end{align*}
where the kernel $\mathcal{K}_{x_1,x_2}$ is given by $\mathcal{K}_{x_1,x_2} = \sum_{j =0}^{V_A -1} \theta_j (x_1) \, \theta_j (x_2)$ \cite{PhysRevB.103.L241118}.
The motivation is to take the thermodynamic limit $V \rightarrow \infty$, and see the dependence of the variance or the standard deviation on $V$. The result is shown in Fig.\ref{fig:dev} (b). Here we have plotted the standard deviation times the total volume with respect to the total volume for two different values of subsystem fraction, namely, $f=1/2$ (purple) and $f=1/4$ (orange). Using the linear-fit, we see that the graph is a straight line with slope $\approx 0.7697$ (purple) and $\approx 0.3820$ (orange) for $f=1/2$ and $f=1/4$ respectively. This suggests that as we increase $V$ (thus approaching the thermodynamic limit), the standard deviation $\braket{\Delta C_A}_G$ will not change, implying that the deviation approaches a constant value which depends on $f$ but is independent of the full system size. This leads us to conjecture that 
\begin{align}
\lim_{V \rightarrow \infty} \braket{\Delta C_A}_G = g(f),
\end{align}
holds for all $f \leq 1/2$, where $g(f)$ is some continuous function of $f$ but is independent of $V$.
This behaviour is similar to case of EE \cite{PhysRevB.103.L241118, PhysRevLett.125.180604} and significantly departs from Page's result, where the variance exponentially vanishes at the thermodynamic limit.

\section{Thermodynamic limit and the Complex SYK$_2$ model}

To  obtain the behaviour of CoE in the the thermodynamic limit, we study the average eigenstate CoE in the complex SYK$_2$ model \cite{PhysRevB.103.104206}, an well-known integrable system. SYK model is a $(0+1)$-dimensional quantum mechanical model of fermions, with all-to-all interactions, with random couplings. It is well-known that SYK$_2$ is integrable whereas SYK$_{q}$ is non-integrable for $q>2$. In recent years, it has emerged as an active research area both from the many-body perspective as well as from the gravity side \cite{Kitaev:1, PhysRevLett.70.3339, Sachdev:2010um, PhysRevD.94.106002, PhysRevB.95.155131}. See \cite{Gu:2017njx, Sonner:2017hxc, PhysRevResearch.2.033505, PhysRevLett.124.106401, Zhang:2020iep, Zhang:2020kia, ish2020sachdevyekitaev} for recent advancements.

Here we consider the complex SYK$_2$ model in Dirac formulation\footnote{In this work, we consider SYK$_2$ Hamiltonian in Dirac formulation. However, one could consider the Majorana formulation. We expect the results do not differ due to particle-hole symmetry \cite{PhysRevB.97.245126, PhysRevB.103.104206}.}. We show that the results we have obtained in previous sections hold for the complex SYK$_2$ model in the thermodynamic limit. The system is given by the following Hamiltonian \cite{PhysRevB.97.245126, PhysRevB.103.104206}
\begin{align}
\mathcal{H} =\sum_{j,k =1}^V M_{j k}\, \hat{c}_j^{\dagger} \hat{c}_k, \label{syk}
\end{align}
where the Hermitian matrix $M$ is constructed from a Gaussian unitary ensemble (GUE). The elements of $M$ are such that entries $M_{i \neq j}$ are complex, whose real and imaginary parts are independent and identically distributed with zero mean and variance $1/V$, whereas $M_{i=j}$, are identically distributed, real entries, with zero mean and variance $2/V$.  The operators $\hat{c}_i^{\dagger}$ and $\hat{c}_j$ are the fermionic creation and annihilation operators at lattice-site $i$ and $j$ respectively. The Hamiltonian \eqref{syk} does not exhibit chaos in the many-body sense, but it is chaotic in the single-body sense \cite{PhysRevB.103.104206}. As observed in \cite{PhysRevB.97.245126}, for this model, the single-body correlation matrix can be obtained by considering the distribution of Jacobi ensemble, which further can be used to calculate entanglement measures. We follow the approach outlined in \cite{PhysRevB.97.245126, PhysRevB.103.104206}. First, we write the capacity for a single many-body eigenstate $\ket{m}$ of the form \cite{Nandy:2021hmk}
\begin{align}
C_A^{(m)} = \sum_{i=1}^{V_A} u_i (1-u_i) \bigg[\ln \bigg(\frac{1-u_i}{u_i}\bigg)\bigg]^2,
\end{align}
where we have restricted to the subsystem $A$, and $u_i$'s are the eigenvalues obtained from the single-body correlation matrix. From the distribution of $\beta =2$-Jacobi ensemble and taking the thermodynamic limit, we get \cite{10.2307/j.ctt7t5vq, PhysRevB.103.104206}
\begin{align}
G_f(u) = \frac{1}{2 \pi f} \frac{\sqrt{u(1-u) + f(1-f) - \frac{1}{4}}}{u(1-u)} \mathbb{I}_{[u_{+},u_{-}]}, \label{dist1}
\end{align}
where $u_{\pm} = \frac{1}{2} \pm \sqrt{f(1-f)}$, and the half-filling condition is imposed. It should be noted that the distribution \eqref{dist1} can be obtained by taking the thermodynamic limit of \eqref{dis} \cite{Ramli}. This allows us to study the behaviour of average CoE in the thermodynamic limit, which can be obtained as $\braket{C_A} = \int d u \, G_f(u) C_A(u)$ with the limits $u \in [u_{-}, u_{+}]$. Note that the distribution \eqref{dist1} vanishes outside this interval. Changing $u = (\xi + 1)/2$, we get the average CoE
\begin{align}\label{capsyk}
\braket{C_A} = \frac{V_A}{4 \pi f} \int_{\xi_{-}}^{\xi_{+}} \mathrm{d \xi} \bigg[\ln \bigg(\frac{1- \xi}{1+\xi}\bigg)\bigg]^2 \sqrt{ f(1-f) - \frac{\xi^2}{4}},
\end{align}
where $\xi_{\pm} = \pm 2 \sqrt{f(1-f)}$ and this expression is valid for $f <1/2$. For $f > 1/2$, we replace $V_A \rightarrow (V - V_A)$ and $f \rightarrow (1- f)$. The above integration can be evaluated exactly in a closed-form for all values of $f$. First, we write the expression \eqref{capsyk} in a series form (see Appendix B for the derivation)
\begin{align}
\braket{C_A} = \frac{ 4 V_{A}}{ \sqrt{\pi}} f(1-f)^{2} \sum_{k = 0}^{\infty}\frac{4^{k} \, \Gamma(k + 3/2) \, \mathcal{H}_{k}}{(k+1)\Gamma(k+3)} \, f^k(1-f)^{k}, \label{analy}
\end{align}
where $\mathcal{H}_k$ is defined as $\mathcal{H}_{k} \equiv H_{2k+2} - \frac{1}{2}H_{k+1}$, and $H_m$ is the $m^{\mathrm{th}}$ Harmonic number. The above series is valid for $f \leq 1/2$, whereas for $f>1/2$, one needs to substitute $f \rightarrow (1-f)$. At $f=1/2$ we can evaluate the sum \eqref{analy} explicitly, which gives $\braket{C_A} = (\pi^2/8 - 1)V_A$.\\

\begin{figure}
	\centering
	\includegraphics[height=5cm,width=0.9\linewidth]{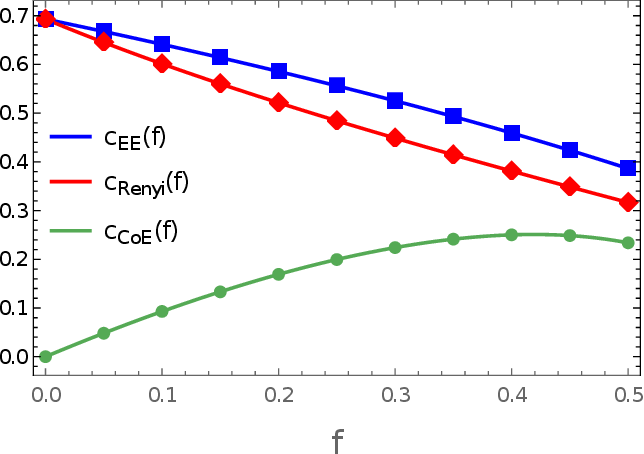}
	\caption{Behaviour of the leading volume-law coefficient of average CoE, EE and second RE with respect to $f$ in complex SYK$_2$ model. The continuous curves are the analytic results. At $f=1/2$, the value of the coefficient of the average EE, second RE and CoE are $2\ln 2-1 ~(\approx 0.3862)$, $2 \ln 2 +2 \ln (2-\sqrt{2}) ~(\approx 0.3167)$ and $\pi^2/8 - 1 ~(\approx 0.2337)$ respectively. The numerical (discrete) plots are obtained for $V=100$, with $100$ Hamiltonian realizations averaged over $10^3$ many-body eigenstates.}
	\label{fig:numsyk}
\end{figure}
It is interesting to note the behaviour of leading volume-law coefficient in Eq.\eqref{analy}. We define the coefficient as
\begin{align}
\lim_{V \rightarrow \infty} \frac{\braket{C_A}}{V_A} = c_{\mathrm{CoE}} (f), \label{defend}
\end{align}
keeping $f=V_A/V$ fixed, and similarly for the average EE and second RE. The behaviour of each coefficient with $f$ are shown in Fig.\ref{fig:numsyk}. Here we note two important results. For $f \rightarrow 0$, the coefficients for the average EE and second RE are $\ln 2$, which corresponds to the maximal average entropy $V_A \ln 2$.  The coefficient of average CoE vanishes as $c_{\mathrm{CoE}}(f\rightarrow 0) =0$, which can be seen from Eq.\eqref{analy}. At $f=1/2$, the average EE and second RE show maximum deviation from their maximal value, but the average CoE is \emph{not} maximum at $f=1/2$. Instead, it reaches a maximal value where the entropies are \emph{not minimal}. This reflects the fact that capacity is a good probe for quantifying partial entanglement structure \cite{PhysRevD.99.066012} which was previously suggested in the context of replica wormholes \cite{Kawabata:2021hac, Okuyama:2021ylc, Kawabata:2021vyo, Nandy:2021hmk}.

\begin{figure}
	\subfigure[]{\includegraphics[height=3cm,width=0.48\linewidth]{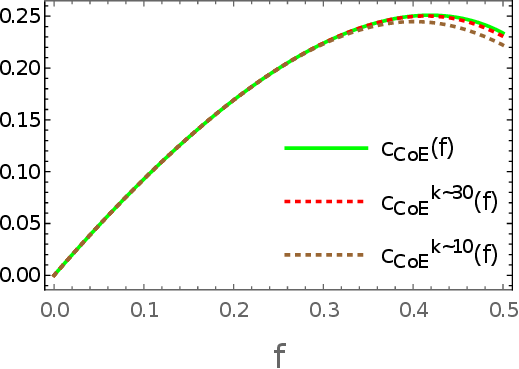}}
	\hfill
	\subfigure[]{\includegraphics[height=3cm,width=0.48\linewidth]{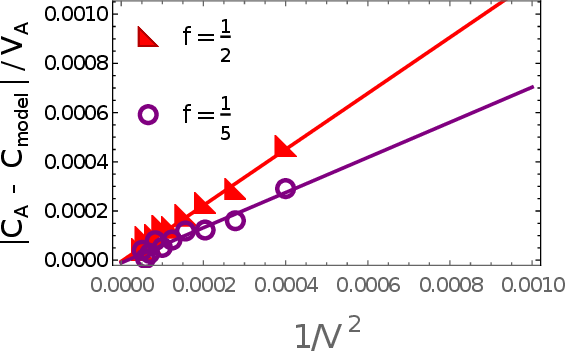}}
	\caption{(a) Convergence of Eq.\eqref{analy} for the leading volume-law coefficient \eqref{defend} of the average CoE. The dotted plots are the results of \eqref{analy} taking the first $k$ terms. (b) Finite-size effect of the coefficient of the average CoE. The fitting has been done for $V \geq 50$. The intercepts ($a_1$) of the linear-fit satisfy $|a_1| \leq 10^{-5}$.} \label{fig:syk}
\end{figure}

To obtain the exact result, one needs to evaluate the infinite sum in Eq.\eqref{analy}. The series can be shown to be convergent (see Appendix C). Especially, we show that at small $f$, the series \eqref{analy} converges exponentially faster than than at $f=1/2$ (see Fig.\ref{fig:syk} (a)). This suggests that considering only first few terms in the series \eqref{analy} provides a good approximation to the exact result. The accuracy and precision depends on $f$, and the number of terms one needs to consider (see Appendix C).

Interestingly, the series \eqref{analy} can be written as a closed-form expression in terms of generalized hypergeometric and Kamp\'e de F\'eriet (KdF) functions
\begin{widetext}
	\begin{align}\label{kdfo}
	\braket{C_A}_G &= 4 f (1-f)^{2} V_{A}\Bigg\{ -\frac{\gamma}{8} \, _3F_2 \left(
	\setlength{\arraycolsep}{0pt}
	\begin{array}{c@{{}~{}}c@{~{}}c}
		1 & 1 & ~\frac{3}{2} \\[1ex]
		~~~2 & 3
	\end{array}
	\;\middle|\;
	4 f(1-f)
	\right) +\frac{(2-\gamma)}{8}\, _3F_2 \left(
	\setlength{\arraycolsep}{0pt}
	\begin{array}{c@{{}~{}}c@{~{}}c}
		1 & 1 & ~\frac{1}{2} \\[1ex]
		~~~2 & 3
	\end{array}
	\;\middle|\;
	4 f(1-f)
	\right) \nonumber \\
	&~~~~~~~~~~~~~~~~~~~~~~~+\frac{\gamma  (3-4 f)}{12(1-f)^2}+ \frac{1}{4}f(1-f)\,F{}^{2:2:1}_{2:1:0}
	\left[
	\setlength{\arraycolsep}{0pt}
	\begin{array}{c@{{}:{}}c@{;{}}c}
		2,\frac{5}{2} & 1,\frac{1}{2} & 1\\[1ex]
		3,4 & \frac{3}{2} & \linefill
	\end{array}
	\;\middle|\;
	4 f(1-f) ,4 f(1-f)
	\right]\Bigg\},
\end{align}
\end{widetext}
where, $\gamma = 0.57721 \cdots$ is the Euler-Mascheroni constant, $_3F_2$ is the generalized hypergeometric function and $F{}^{2:2:1}_{2:1:0}$ is the KdF function. See Appendix D for the derivation. However, we prefer to use the simpler form \eqref{analy} here to compare with numerical results.

As an alternative justification, one can directly check the result at $f=1/2$ using the replica method. Using the expression for RE \cite{Zhang:2020kia} and the definition \eqref{capdef}, we obtain the leading volume-law coefficient of the average CoE at $f=1/2$ as
\begin{align}
c_{\mathrm{CoE}}\Big|_{f=\frac{1}{2}} = \frac{4}{\pi}\lim_{n \rightarrow 1} n^2 \partial_n^2 \int_{0}^1  \mathrm{d k}  \, \frac{\ln (1+k^{2n})}{1 + k^2}  = \frac{\pi^2}{8} - 1,
\end{align}
which confirms our result.

We numerically \cite{Peschel_2003, Peschel_2009, xiao2009theory} calculate the coefficients of the average CoE, EE and second RE for $V=100$, with $100$ Hamiltonian realizations and averaging over $10^3$ many-body eigenstates. We compare the numerical results with the analytic result in Eq.\eqref{analy} in the thermodynamic limit (see Fig.\ref{fig:numsyk}). It should be noted that the analytic result is in the thermodynamic limit, so a small deviation due to the finite-size is expected. We show the deviation due to the finite-size in Fig.\ref{fig:syk} (b) (also see Appendix E). Using the linear-fit ($a_0/V^2 + a_1)$, we observe that the intercepts of the straight lines satisfy $|a_1| \leq 10^{-5}$. This implies that the finite-size effects are small and become negligible as we approach the thermodynamic limit.

Thus, we see that the behaviour of the average CoE is very different from that of the average EE and second RE. This is also apparent in the small $f$ limit, where the entropies (density) are concave functions of $f$, whereas the average CoE (density) is a convex function of $f$. This can also be observed from the series expression \eqref{analy}. After a small $f$ expansion and taking the first few terms of the series, we get 
\begin{align}
\frac{d^2 A(f)}{d f^2}\bigg|_{f \rightarrow 0} >~ 0,
\end{align}
where $A(f) \equiv \braket{C_{A}}/(V \ln 2)$ i.e., the average CoE is convex at small $f$. One can also see that $d^{2}A(f)/df^{2}<0$ near $f=1/2$, i.e., it becomes concave (see Fig.\ref{fig:page}). This property is unique to the capacity and is absent in the entanglement or R\'enyi entropies.

One can also note that while the CoE and its variance are continuous functions of the subsystem fraction $f$, they are not necessarily analytic in nature. It can be easily seen (in the thermodynamic limit) that the derivative of \eqref{analy} has a sign discontinuity at $f = 1/2$. Specifically, we have the result
\begin{equation}
	\frac{d}{d f}\braket{C_{A}}|_{f = \frac{1}{2}} = \begin{cases}
		- \frac{4}{\sqrt{\pi}}\sum_{k = 0}^{\infty}\frac{\Gamma(k + \frac{3}{2}) \mathcal{H}_{k}}{4(k+1)\Gamma(k + 3)} &\; f \rightarrow \frac{1}{2}-\\
		\frac{4}{\sqrt{\pi}}\sum_{k = 0}^{\infty}\frac{\Gamma(k + \frac{3}{2}) \mathcal{H}_{k}}{4(k+1)\Gamma(k + 3)} &\; f \rightarrow \frac{1}{2}+
	\end{cases}
\end{equation}
Similarly, we expect infinite divergences in higher derivatives. Therefore, we do not expect the CoE or its variance to be analytic functions even for finite-size systems, although we have not explicitly demonstrated that in the general case.

It is interesting to compare the results of CoE for random pure states and the results we have obtained here.  The computation of capacity of entanglement has been done very recently for random pure states (which is exactly similar in the spirit to Page's original calculation) in \cite{Okuyama:2021ylc}. It is easy to check that at $f \rightarrow 0$, the capacity indeed vanishes for random pure states. This coincides with our result and consistent with our findings, at vanishing subsystem fraction one cannot distinguish between the average CoE for random pure state and Gaussian states that we have considered in this paper. This is due to the fact that at this limit, the spectra of reduced density matrices for both cases behave in similar way.

This motivates us to see whether the average eigenstate CoE, having different characteristic features as compared to the average eigenstate entropies, can distinguish between integrable and quantum-chaotic systems \cite{PhysRevE.99.032111}. It has been shown that in the case of quantum-chaotic Hamiltonian, the average eigenstate EE is near to maximal and closely follows Page's result \cite{PhysRevLett.119.220603, PhysRevE.99.032111}, which shows deviation near $f=1/2$, only at finite temperature \cite{PhysRevE.100.022131}. This suggests that capacity indeed vanishes for the quantum-chaotic systems. On the other hand, our result Eq.\eqref{kdfo} indicates that the volume-law coefficient is non-vanishing for finite subsystem fraction. Also note at vanishing subsystem fraction, even for the integrable system, average CoE vanishes. This justifies the fact that for small subsystem size, the distinguishability between integrability and non-integrability breaks down. Hence, to see the difference, one needs to consider finite $f$, say at $f=1/2$.  This suggests that for finite subsystem size, up to the leading order, non-vanishing  volume-law coefficient of average CoE might indicate that the system is integrable, whereas vanishing coefficient implies the system is quantum-chaotic. Thus, similar to the average EE, the average eigenstate CoE is an excellent probe for distinguishing between integrable and quantum-chaotic systems.


\section{Conclusion and Summary}

In this paper, we have studied the average eigenstate capacity of entanglement in fermionic Gaussian states and obtained explicit analytic expressions. We obtain the analytic expressions for average CoE for both finite-size systems and in the thermodynamic limit. In thermodynamic limit, we further express it in terms of generalized hypergeometric and Kamp\'e de F\'eriet (KdF) functions. However, we could not find a closed-form expression of average CoE for finite-size systems. 
The average CoE possesses different properties as compared to the average entropies; for example, it is a convex function for small subsystem size, but it starts to become concave once the subsystem size increases. In contrast to the entropies, it proportionally depends on subsystem fraction. Moreover, based on numerical findings, we conjecture the variance of the average CoE to be a  continuous function of $f$. We consider the variance numerically and it would be very interesting to find an analytic, closed-form expression of it.

We have considered the complex SYK$_2$ model, both analytically and numerically, in the thermodynamic limit. We have derived an analytic and closed-form expression of the average CoE which we have verified numerically. As observed in Fig.\ref{fig:numsyk}, the average CoE is not maximum at $f=1/2$, which is attributed to the fact that the CoE is a good probe for the partial entanglement structure \cite{PhysRevD.99.066012}. It will be interesting to understand the mechanism behind this behaviour in terms of entanglement. We have also shown in Fig.\ref{fig:syk} (b) that the finite-size effects are negligible as we approach the thermodynamic limit. Finally, based on above findings, we propose the average eigenstate CoE as a useful probe to distinguish between integrable and quantum-chaotic systems.

One can study the CoE for canonical thermal pure quantum (cTPQ) states \cite{PhysRevLett.111.010401, PhysRevLett.108.240401, PhysRevE.99.032111, Fujita:2018wtr}, which might play an important role in the context of the black hole information problem, as CoE has already been established as a good probe in identifying the Page time \cite{Kawabata:2021hac, Kawabata:2021vyo}. Also, as a natural extension, it would be very exciting to investigate the CoE at finite temperature, and to see its effect on the nature of the Page curve itself. Moreover, as pointed out in \cite{PhysRevE.100.022131}, near $f=1/2$ for quantum-chaotic Hamiltonians, the subleading correction to the EE is proportional to the square root of heat capacity. It would be interesting to obtain the subleading corrections to the CoE near $f=1/2$.

\section{Acknowledgments}
	We thank B. Ananthanarayan, Aranya Bhattacharya, Arpan Bhattacharyya, Pawel Caputa, Chethan Krishnan and Aninda Sinha for useful suggestions and critical comments on the draft. We would like to thank the anonymous referees for useful comments and suggestions. BB is supported by the Ministry of Human Resource Development (MHRD), Government of India through the Prime Ministers' Research Fellowship. PN acknowledges financial support from University Grants Commission (UGC), Government of India.

\appendix
\section{ Appendix A: Derivation of Eq.(\ref{pandora})}

In this section, we derive Eq.\eqref{pandora}. First, we rewrite the capacity \eqref{cmain} in the following form
\begin{align*}
	c(x) = \frac{1}{4}(1-x^2) \, \lim_{\epsilon \rightarrow 0} \partial_{\epsilon}^2 \, \bigg(\frac{1+x}{1-x}\bigg)^{\epsilon}. \label{capnew}
\end{align*}
To calculate the average capacity, we consider the integral (Eq.\eqref{cc1})
\begin{equation}
	I_{\epsilon}=  \frac{1}{8}\int_{-1}^{1}\mathrm{d x} \, (1-x^2)  \, \bigg(\frac{1+x}{1-x}\bigg)^{\epsilon}  \rho(|x|), \label{app1}
\end{equation}
with the density function in Eq.\eqref{dis} and corresponding coefficients. To perform the above integral, we consider the following series representation of Jacobi Polynomial
\begin{align*}
	P_{n}^{(\alpha, \beta)}(z)=&\frac{\Gamma(\alpha+n+1)}{\Gamma(n+1) \Gamma(\alpha+\beta+n+1)} \nonumber \\ & \times \sum_{m=0}^{n}\binom{n}{m} \frac{\Gamma(\alpha+\beta+n+m+1)}{\Gamma(\alpha+m+1)}\Big(\frac{z-1}{2}\Big)^{m}
\end{align*}
Substituting the above expression in Eq.\eqref{app1} we get 
\begin{widetext}
	\begin{align}
		I_{\epsilon}=\sum_{j=0}^{V_A-1}\sum_{k,m=0}^{2j}\frac{(2 \Delta +4 j+1) (-1)^{k+m} \binom{2 j}{k} \binom{2 j}{m} 2^{-2 \Delta -k-m-2}  \Gamma (2 j+k+2 \Delta +1) \Gamma (2 j+m+2 \Delta +1)}{\Gamma (2 j+1) \Gamma (2 j+2 \Delta +1) \Gamma (k+\Delta +1) \Gamma (m+\Delta +1)V_{A}}\nonumber \\ 
		\times\int_{-1}^{-1}\mathrm{d x} \, (x+1)^{\Delta +\epsilon +1}(1-x)^{\Delta +k+m-\epsilon +1}.
	\end{align}
	We use the following integral
	\begin{align*}
		\int_{-1}^1 \mathrm{d x} \, (1-x)^a (x+1)^b =\frac{2^{a+b+1} \Gamma (a+1) \Gamma (b+1)}{\Gamma (a+b+2)},
	\end{align*}
	to get 
	\begin{align}
		I_{\epsilon}=\sum_{j=0}^{V_A-1}\sum_{m,k=0}^{2j}\frac{ (-1)^{k+m} \binom{2 j}{k} \binom{2 j}{m}  \Gamma (2 j+k+2 \Delta +1) \Gamma (2 j+m+2 \Delta +1) \Gamma (k+m+\Delta -\epsilon +2)}{V_{A}\Gamma (2 j+1) \Gamma (2 j+2 \Delta +1) \Gamma (k+\Delta +1) \Gamma (m+\Delta +1) \Gamma (k+m+2 \Delta +4)}\\ \nonumber 
		\times (2 \Delta +4 j+1)\Gamma (\Delta +\epsilon +2).
	\end{align}
\end{widetext}
Differentiating twice with respect to $\epsilon$ and taking the limit $\epsilon \rightarrow 0$, we get Eq.\eqref{pandora}.

\section{ Appendix B: Derivation of Eq.(\ref{analy}) }

In order to derive Eq.\eqref{analy}, we use the following series expansion
\begin{equation*}
	\ln \bigg( \frac{1-x}{1+x} \bigg) = -2\sum_{n=0}^{\infty}\frac{x^{2n+1}}{2n+1},
\end{equation*}
valid for $|x|\leq 1$, which one can see holds for $\xi$. Therefore, the integral \eqref{capsyk} becomes 
\begin{align*}
	\braket{C_A} = \frac{V_{A}}{4 \pi f} \int_{\xi_{-}}^{\xi_{+}} \mathrm{d \xi} \, \sum_{m = 0}^{\infty} \sum_{n=0}^{\infty}\frac{4\xi^{2m + 2n + 2}}{(2m+1)(2n+1)}\\
	\times \sqrt{f(1-f) - \frac{\xi^{2}}{4}}.
\end{align*}
Transforming the summation indices $(m,n) \mapsto (k,n)$, where $k=m+n$, we obtain the following summation
\begin{align*}
	\braket{C_A} = \frac{V_{A}}{4 \pi f} \int_{\xi_{-}}^{\xi_{+}} \mathrm{d \xi} \, \sum_{k = 0}^{\infty} \sum_{n=0}^{k}\frac{4\xi^{2k + 2}}{(2k- 2n +1)(2n+1)}\\ \nonumber
	\times \sqrt{f(1-f) - \frac{\xi^{2}}{4}}.
\end{align*}
Rearranging the above equation, we can write it as
\begin{align*}
	\braket{C_A} = \frac{V_{A}}{4 \pi f}&\, \sum_{k = 0}^{\infty}\bigg( \sum_{n=0}^{k}\frac{1}{(2k- 2n +1)(2n+1)} \bigg)\\ \nonumber
	&\times \int_{\xi_{-}}^{\xi_{+}} \mathrm{d \xi} 4\xi^{2k + 2}\sqrt{f(1-f) - \frac{\xi^{2}}{4}}.
\end{align*}
The summation in the parenthesis can be evaluated as follows.
\begin{align*}
	\sum_{n=0}^{k} \frac{1}{(2k- 2n +1)(2n+1)} &= \frac{1}{k+1}\sum_{n=0}^{k} \frac{1}{2n+1} \\ \nonumber
	&= \frac{2 H_{2k+2} - H_{k+1}}{2 k+2} \equiv \frac{\mathcal{H}_{k}}{k+1},
\end{align*}
where we have used the definition
\begin{equation*}
	\mathcal{H}_{k} \equiv  H_{2k+2} - \frac{1}{2}H_{k+1}. \label{har1}
\end{equation*}
Here $H_{p}$ denotes the $p^{\text{th}}$ harmonic number. Thus, we get
\begin{equation}\label{taft}
	\braket{C_A} = \frac{V_{A}}{4 \pi f} \sum_{k = 0}^{\infty}\frac{4 \mathcal{H}_{k}}{k+1 }\int_{\xi_{-}}^{\xi_{+}} \mathrm{d \xi} \, \xi^{2k + 2}\sqrt{f(1-f) - \frac{\xi^{2}}{4}}
\end{equation} 
Now, this integral is evaluated to get the following result
\begin{align*}
	\int_{\xi_{-}}^{\xi_{+}} d \xi \, \xi^{2k + 2}\sqrt{f(1-f) - \frac{\xi^{2}}{4}} &= \frac{2^{2k+2} (f(1-f))^{k + 2}}{\Gamma(k+3)}\\ \nonumber
	&~~~~~~~\times \, \Gamma(k + 3/2)\sqrt{\pi}.
\end{align*}
Plugging this result in Eq.\eqref{taft} gives Eq.\eqref{analy}.

\section{ Appendix C: Convergence of the series (\ref{analy})}\label{conv}
In this section, we examine the convergence properties of Eq.\eqref{analy}. The series can be shown to be convergent by using the Ratio test. From Eq.\eqref{analy}, we have the $n^\mathrm{th}$ term of the series Eq.\eqref{analy}
\begin{equation*}
	a_{n}= \frac{4^n f^n (1-f)^n  \left(H_{2 n+2}-\frac{1}{2} H_{n+1}\right) \Gamma \left(n+3/2\right)}{(n+1) \Gamma (n+3)}.
\end{equation*}
We consider the following limit
\begin{equation}
	L=\lim_{n \rightarrow \infty}\left|\frac{a_{n+1}}{a_{n}}\right|= 4f(1-f).
\end{equation}
Hence, we have $L<1$ for $f<1/2$, and the series is convergent. For $f=1/2$, we have $L= 1$, and thus the test is inconclusive. Therefore, we use Raabe's test to check for the convergence. We consider the following
\begin{equation}
	L= \lim_{n \rightarrow \infty}n\left(\frac{a_{n}}{a_{n+1}}-1\right)= \frac{5}{2}.
\end{equation}
Since $L>1$, we conclude that the series converges at $f=1/2$, via Raabe's test.

It is also interesting to look at the convergence rate of the series at $f=1/2$ and for  $f<1/2$. To do this, we note the ratio of the $n^{\mathrm{th}}$ term of the series for $f<1/2$ and at $f=1/2$, which is given by
\begin{equation}
	R= \frac{a_n|_{f<1/2}}{a_n|_{f=1/2}}= 4^{n} f^{n} (1-f)^{n}. \label{rate}
\end{equation}
By comparing the $n^{\mathrm{th}}$ term of the series in the above two cases, we see that for $f<1/2$, the terms are suppressed exponentially in comparison to the terms at $f=1/2$. Hence, the series converges exponentially faster for $f<1/2$ than at $f=1/2$ (see Fig.\ref{fig:syk} (a)). This can be further seen by the observation that, up to a required accuracy, the evaluation of the series for $f<1/2$  takes significantly less number of terms to get the desired result as compared $f=1/2$. As an example, for $f=1/3$, evaluating $60$ terms in the series \eqref{analy} gives a result that matches up to $6$ significant digits with the result from numerical integration of Eq.\eqref{capsyk}. On the other hand, for $f=1/2$, the evaluation of the first $1000$ terms give a result which matches only up to $3$ significant digits with numerical integration result. Hence, for small $f$, the sum \eqref{analy} is equally good as compared with the exact result.

\section{ Appendix D: Closed form of Eq.(\ref{analy})}
Here we render Eq.\eqref{analy} in a closed-form expression. From Eq.\eqref{analy}, we consider the sum
\begin{equation}
	S =\sum_{k = 0}^{\infty}\frac{4^{k} \, \Gamma(k + 3/2) \, \mathcal{H}_{k}}{(k+1)\Gamma(k+3)} \, f^k(1-f)^{k}, \label{sumsyk}
\end{equation}
where  $\mathcal{H}_{k} \equiv H_{2k+2} - \frac{1}{2}H_{k+1}$. We note the following identities for the digamma function
\begin{align*}
	\Psi(z+1)&= \Psi(z)+ \frac{1}{z},\\
	2 \, \Psi(2z)&= 2\ln 2+ \Psi(z)+ \Psi\Big(z+\frac{1}{2}\Big),
\end{align*}
where the first one is the recurrence relation and the second is the Legendre duplication formula for digamma functions. Using the relations above we can write
\begin{equation}
	\mathcal{H}_{k} = \frac{1}{2}\Psi\Big(z+\frac{1}{2}\Big)+ \ln 2 + \frac{2 \gamma  k+\gamma +2}{4 k+2},
\end{equation}
where $\gamma$ is the Euler-Mascheroni constant. Substituting the above in Eq.\eqref{sumsyk}, we get
\begin{align}\label{sumsyk2}
	S =\sum_{k = 0}^{\infty}\frac{4^{k} \, \Gamma(k + \frac{3}{2}) \, \big(\frac{1}{2}\Psi (k+\frac{1}{2})+ \ln 2 + \frac{2 \gamma  k+\gamma +2}{4 k+2}\big)}{(k+1)\Gamma(k+3)} \,\\ \nonumber
	\times ~f^k(1-f)^{k}.
\end{align}
Of the three terms in the above summation the second and the third can be evaluated using MATHEMATICA. We get the following
\begin{align}
	\sum_{k = 0}^{\infty}\frac{4^{k} \, \Gamma(k + \frac{3}{2}) \, \ln 2}{(k+1)\Gamma(k+3)} \, & f^k(1-f)^{k}= \frac{\ln 2}{4} \nonumber \\ 
	&\times~  _3F_2 \left(
	\setlength{\arraycolsep}{0pt}
	\begin{array}{c@{{}~{}}c@{~{}}c}
		1 & 1 & ~\frac{3}{2} \\[1ex]
		~~~2 & 3
	\end{array}
	\;\middle|\;
	4 f(1-f)
	\right), \label{s2} 
\end{align}
and
\begin{align*}
	\sum_{k = 0}^{\infty}\frac{4^{k} \, \Gamma(k + \frac{3}{2}) \, \big(\frac{2 \gamma  k+\gamma +2}{4 k+2}\big)}{(k+1)\Gamma(k+3)} \, f^k(1-f)^{k}=\frac{\gamma  (6-8 f)}{24(f-1)^2}\\ \nonumber
	+\frac{(2-\gamma)}{8} \, ~ _3F_2 \left(
	\setlength{\arraycolsep}{0pt}
	\begin{array}{c@{{}~{}}c@{~{}}c}
		1 & 1 & ~\frac{1}{2} \\[1ex]
		~~~2 & 3
	\end{array}
	\;\middle|\;
	4 f(1-f)
	\right). \label{s3}
\end{align*}
The first part in Eq. \eqref{sumsyk2} seems intractable at first due to the presence of $\Psi(k+1/2)$ function so we cannot sum it directly. To do it we use a trick similar to the Feynman's trick for integrals, but for the summation here. We notice that if the summand is only the ratio of gamma functions  then the sum can be written in terms of generalized hypergeometric functions. We further note that
\begin{align*}
	\frac{\mathrm{d} \,\Gamma(x)}{\mathrm{d x}}= \Psi(x)\Gamma(x).
\end{align*}
Denoting the sum by $S_1$ we can write following
\begin{align}
	S_{1}= \sum_{k = 0}^{\infty}\frac{4^{k} \Gamma(k + \tfrac{3}{2}) }{2(k+1)\Gamma(k+3)} f^k(1-f)^{k} \nonumber \partial_{z}\left(\frac{\Gamma(k+z+\tfrac{1}{2})}{\Gamma(k+\tfrac{1}{2})}\right)\Bigg|_{z=0}.
\end{align}
The derivative with respect to $z$ in $S_1$ can be extracted out and the remaining series summation over gamma functions can be summed to give
\begin{align*}
	S_{1}=\partial_{z}\left[\frac{\Gamma \left(z+\tfrac{1}{2}\right)}{8 \sqrt{\pi }} ~  _4F_3 \left(
	\begin{array}{c@{{}~{}}c@{~{}}c@{~{}}c@{~{}}}
		1 & 1 & \frac{3}{2} & z+\tfrac{1}{2} \\[1ex]
		~\frac{1}{2} & ~2 & ~3
	\end{array}
	\;\middle|\;
	4 f(1-f)
	\right)\right]_{z=0}.
\end{align*}
Using the chain rule we perform the derivative and we get the following
\begin{align}
	S_{1}&= \frac{\Psi(1/2)}{8}\,  _3F_2 \left(
	\setlength{\arraycolsep}{0pt}
	\begin{array}{c@{{}}c@{{}}c@{{}}c@{{}}}
		1 & 1 & ~\frac{3}{2} \\[1ex]
		~~2 & ~~3
	\end{array}
	\;\middle|\;
	4 f(1-f)
	\right)\\ \nonumber
	&+\frac{1}{8}\,\partial_{(z+\frac{1}{2})}\left[_4F_3 \left(
	\setlength{\arraycolsep}{0pt}
	\begin{array}{c@{{} {}}c@{ {}}c@{ {}}c@{ {}}}
		1 & 1 & \frac{3}{2} & z+\frac{1}{2} \\[1ex]
		~\frac{1}{2} & ~2 & ~3
	\end{array}
	\;\middle|\;
	4 f(1-f)
	\right)\right]_{z=0},
\end{align}
where we have used the chain rule to perform the derivative. This is the same as taking the derivative of the above $_4F_3$ with respect to one of its parameters. The derivatives of the generalized hypergeometric function with respect to its parameters can be written in term of KdF function (Eq.\eqref{kdfdef}). Here we quote the result 
\begin{widetext}
	\begin{align}
		\frac{\partial}{\partial a_{4}}\left[\,  _4F_3 \left(
		\setlength{\arraycolsep}{0pt}
		\begin{array}{c@{{}~{}}c@{~{}}c@{~{}}c@{~{}}}
			a_1 & a_2 & a_3 & a_4 \\[1ex]
			~~~~b_1 & ~b_2 & ~b_3
		\end{array}
		\;\middle|\;
		x
		\right)\right]= x\frac{a_{1}a_{2}a_{3}}{b_{1}b_{2}b_{3}} \, F^{4:2:1}_{4:1:0}
		\left[
		\setlength{\arraycolsep}{0pt}
		\begin{array}{c@{{}:{}}c@{;{}}c}
			a_{1}+1,a_{2}+1,a_{3}+1,a_{4}+1 & 1,a_{4} & 1\\[1ex]
			b_{1}+1,b_{2}+1,b_{3}+1,2 & a_{4}+1 & \linefill
		\end{array}
		\;\middle|\;
		x,x
		\right],
	\end{align}
	where $F^{4:2:1}_{4:1:0}$ is the KdF function. Using the above relation, we get
	\begin{align}
		S_{1}&= \frac{\Psi(1/2)}{8}\,  _3F_2 \left(
		\setlength{\arraycolsep}{0pt}
		\begin{array}{c@{{}}c@{{}}c@{{}}c@{{}}}
			1 & 1 & ~\frac{3}{2} \\[1ex]
			~~~2 & ~~3
		\end{array}
		\;\middle|\;
		4 f(1-f)
		\right)+
		\frac{1}{4}f(1-f) \, F{}^{2:2:1}_{2:1:0}
		\left[
		\setlength{\arraycolsep}{0pt}
		\begin{array}{c@{{}:{}}c@{;{}}c}
			2,\frac{5}{2} & 1,\frac{1}{2} & 1\\[1ex]
			3,4 & \frac{3}{2} & \linefill
		\end{array}
		\;\middle|\;
		4 f (1-f),4 f(1-f)
		\right].
	\end{align}
	Substituting all the results in Eq.\eqref{sumsyk2}, we get 
	\begin{align}\label{kdf}
		\braket{C_A}_G &= 4 f (1-f)^{2} V_{A}\Bigg\{ -\frac{\gamma}{8} \, _3F_2 \left(
		\setlength{\arraycolsep}{0pt}
		\begin{array}{c@{{}~{}}c@{~{}}c}
			1 & 1 & ~\frac{3}{2} \\[1ex]
			~~~2 & 3
		\end{array}
		\;\middle|\;
		4 f(1-f)
		\right) +\frac{(2-\gamma)}{8}\, _3F_2 \left(
		\setlength{\arraycolsep}{0pt}
		\begin{array}{c@{{}~{}}c@{~{}}c}
			1 & 1 & ~\frac{1}{2} \\[1ex]
			~~~2 & 3
		\end{array}
		\;\middle|\;
		4 f(1-f)
		\right) \nonumber \\
		&~~~~~~~~~~~~~~~~~~~~~~~+\frac{\gamma  (3-4 f)}{12(1-f)^2}+ \frac{1}{4}f(1-f)\,F{}^{2:2:1}_{2:1:0}
		\left[
		\setlength{\arraycolsep}{0pt}
		\begin{array}{c@{{}:{}}c@{;{}}c}
			2,\frac{5}{2} & 1,\frac{1}{2} & 1\\[1ex]
			3,4 & \frac{3}{2} & \linefill
		\end{array}
		\;\middle|\;
		4 f(1-f) ,4 f(1-f)
		\right]\Bigg\},
	\end{align}
\end{widetext}
where, $\gamma = 0.57721 \cdots$ is the Euler-Mascheroni constant, $_3F_2$ is the generalized hypergeometric function and $F{}^{2:2:1}_{2:1:0}$ is the KdF function (we have omitted the arguments for brevity). It should be stressed that the above expression is in the thermodynamic limit. We mention that the $_3F_2$ functions can be further written in terms of simpler functions using MATHEMATICA command \texttt{FunctionExpand}, but the form given here is more compact and thus we leave them in this form.

Next, we discuss the convergence properties of the KdF function that we encountered in Eq.\eqref{kdf}. The KdF functions are two variable generalizations of hypergeometric function. They are used to represent the derivative of generalized hypergeometric functions, $_pF_q$, with respect to one of its parameters. They can also arise in the analytic continuation formulae for 2-variable Horn's function; for example, they arise in the standard analytic continuation of Appell $F_2$ around $(\infty,\infty)$. They also appear in the indefinite integrals of the products of two and three Meijer G-functions. The KdF function is defined as follows
\begin{widetext}
	\begin{align}
		F{}^{p:q:k}_{l:m:n}
		\left[
		\setlength{\arraycolsep}{0pt}
		\begin{array}{c@{{}:{}}c@{;{}}c}
			(a_p) & (b_q) & (c_k)\\[1ex]
			(\alpha_l) & (\beta_m) & (\gamma_n)
		\end{array}
		\;\middle|\;
		x,y
		\right]=
		\sum_{r=0}^{\infty}\sum_{s=0}^{\infty}\frac{\prod_{j=1}^{p}(a_j)_{r+s}\prod_{j=1}^{q}(b_j)_{r}\prod_{j=1}^{k}(c_j)_{s}}{\prod_{j=1}^{l}(\alpha_j)_{r+s}\prod_{j=1}^{m}(\beta_j)_{r}\prod_{j=1}^{n}(\gamma_j)_{s}}\frac{x^r}{r!}\frac{y^s}{s!}. \label{kdfdef}
	\end{align} 
\end{widetext}
Unlike the one variable case where the radius of convergence is straightforward to calculate, the region of convergence (ROC) determination for the two-variable case is a non-trivial task. This is achieved using Horn's theorem. For the case of the KdF function the following standard results are obtained in \cite{Srivastava, PhysRevD.101.116008}
\begin{enumerate}
	\item If $p+q<1+m+1$, ~$p+k<l+n+1$, then ROC is $|x|<\infty$, $|y|<\infty$,
	\item If $p+q=l+m+1$, ~$p+k=l+n+1$, then the ROC is given by 
	\begin{align}
		\left\{
		\begin{array}{ll}
			|x|^{\frac{1}{p-l}}+|y|^{\frac{1}{p-l}}   & ~~\mbox{if } ~p> l \\
			\text{max}{|x|,,|y|}<1 & ~~\mbox{if } ~p \leq l
		\end{array}
		\right. \label{kdfconv}
	\end{align}
\end{enumerate}
An important point to note is that the above convergence conditions do not tell anything about the boundary of the ROC. Also, since the KdF function has to be evaluated as a double sum, so it is imperative to check how many terms one needs to add in the series to get a precise result for a given $(x,y)$. The convergence rate for the double series of KdF is dependent on how close the point is to the boundary of the ROC. The closer the point is, the more the number of terms that are required to be added in order to get the result within a certain degree of accuracy. From  Eq.\eqref{kdf} we get the KdF as $F^{2:2:1}_{2:1:0}\, \big[4 f(1-f),4 f(1-f) \big]$, suppressing other arguments. We see that it satisfies the second condition in \eqref{kdfconv}. Since $p=l$, and both $x$ and $y$ are same for our case, the ROC of the KdF function leads to the condition
\begin{align}
	\text{max}(4 f(1-f))<1
\end{align} 
which holds for $f<1/2$. For $f=1/2$, Eq.\eqref{analy} is explicitly summed over, and we get $\braket{C_A} = (\pi^2/8-1)V_A$.

\vspace{1cm}

\section{Appendix E: Numerical calculation}\label{numerical}
For numerical evaluation, we use the method outlined in \cite{PhysRevLett.125.180604, PhysRevB.103.104206, Peschel_2003, Peschel_2009}. For the SYK$_2$ model, we have the following quadratic Hamiltonian
\begin{align}
	\mathcal{H} =\sum_{j,k =1}^V M_{j k}\, \hat{c}_j^{\dagger} \hat{c}_k,
\end{align}
The above can be diagonalized \cite{xiao2009theory} to give the following form
\begin{align}
	\mathcal{H} =\sum_{a=1}^{V} \varepsilon_{a}\, \hat{d}_{a}^{\dagger} \hat{d}_{a},
\end{align}
where $\varepsilon$ is the diagonal matrix.
The $m$-body eignekets for $\mathcal{H}$ can be written as \cite{PhysRevB.103.104206}
\begin{align}
	\ket{m}= \prod_{{\{p_{l}\}}_{m}}\hat{c}_{l}^{\dagger} \ket{0},  \quad m=1,2,3,\cdots,2^{V}.
\end{align}
where $\{p_{l}\}_{m}$ denotes the $m^{\mathrm{th}}$ set of  occupied one particle eigenket. We then introduce the number operator $N_{p}= 2(\hat{d}_a^{\dagger} \hat{d}_a)-1$ for which $N_{p} \ket{m}= N_{p}^{m}\ket{m}$, $N_{p}^{m}= 1(-1)$ for site being filled (empty). Let $U$ be the diagonalizing matrix for $M$, such that $\hat{c}_{i}= \sum_{q=1}^{V}U_{ia}\hat{d}_{i}$. Then the one body correlation matrix is given by \cite{PhysRevB.103.104206}
\begin{align}
	(\mathcal{J}_{m})_{ij}= 2 \braket{ m| \hat{c}_i^{\dagger} \hat{c}_j|m }+ \delta_{ij}= \sum_{p=1}^{V}N_{p}^{m}\hat{U}^{*}_{ip}\hat{U}_{jp},
\end{align}
where $i,j \leq V_{A}$. Let $\lambda_{i}; ~i=1,2,\cdots,V_{A} $ be the eigenvalues of the correlation matrix $\mathcal{J}_{m}$. Then we can write the capacity as 
\begin{equation}
	C_{m}= \frac{1}{4} \sum_{i=1}^{V_A}\left(1-\lambda _i^2\right) \bigg[\log \left(\frac{1+\lambda _i}{1-\lambda _i}\right)\bigg]^2.
\end{equation}
Averaging over all eigenstates, the average CoE is then given by 
\begin{equation}\label{capnum}
	\overline{C}= \frac{1}{2^{V}}\sum_{m=1}^{2^{V}}C_{m}.
\end{equation}
We calculate the value of the average eigenstate CoE using the analytic expression Eq.\eqref{analy} and numerically using Eq.\eqref{capnum}. We take $V=100$, with $100$ Hamiltonian realizations and averaged over $10^3$ many-body eigenstates. The numerical plots agree with the analytical result (see Fig.\ref{fig:syk} (a)).

As the analytical result is valid in the thermodynamic limit, whereas the numerical results are obtained by considering finite lattice size, it is instructive to analyze the finite-size correction. For this, we calculate the following \emph{deficit}
\begin{equation}
	\mathcal{D} \equiv \frac{|C_{\mathrm{analytic}}-\overline{C}_{\mathrm{model}}|}{V_{A}}.
\end{equation}
\begin{widetext}
	The table  shows the numerical value of $\mathcal{D}$ for $f=1/5, 1/2$ and $V = 50,60,\cdots,140$.
	\begin{table}[!h]
		\centering
		\resizebox{\columnwidth}{!}{
			\begin{tabular}{|l|l|l|l|l|l|l|l|l|l|l|}
				\hline
				\multicolumn{1}{|c|}{\textbf{f\textbackslash{}V}} & \multicolumn{1}{c|}{\textbf{$50$}} & \multicolumn{1}{c|}{\textbf{$60$}} & \multicolumn{1}{c|}{\textbf{$70$}} & \multicolumn{1}{c|}{\textbf{$80$}} & \multicolumn{1}{c|}{\textbf{$90$}} & \multicolumn{1}{c|}{\textbf{$100$}} & \multicolumn{1}{c|}{\textbf{$110$}} & \multicolumn{1}{c|}{\textbf{$120 $}} & \multicolumn{1}{c|}{\textbf{$130$}} & \multicolumn{1}{c|}{\textbf{$140$}} \\ \hline
				$1/5 $                                    & $2.91\times10^{-4}$                     & $1.60\times10^{-4}$                     & $1.23\times10^{-4}$                     & $1.18\times10^{-4}$                     & $8.17\times10^{-5}$                     & $5.21\times10^{-5}$                      & $7.93\times10^{-5}$                      & $2.76\times10^{-5}$                       & $6.56\times10^{-6}$                   & $4.01\times10^{-5}$                      \\ \hline
				$1/2 $                                    & $4.63\times10^{-4}$                     & $2.89\times10^{-4}$                     & $2.31\times10^{-4}$                     & $1.81\times10^{-4}$                     & $1.32\times10^{-4}$                     & $1.38\times10^{-4}$                      & $1.00\times10^{-4}$                      & $6.79\times10^{-5}$                       & $9.16\times10^{-5}$                      & $4.24\times10^{-5}$                      \\ \hline
		\end{tabular}}
	\end{table}
\end{widetext}
The above values are plotted against $1/V^2$. We see the deficit vanishes as we increase $V$, i.e., as we approach the thermodynamic limit. The following are obtained using linear-fit (see Fig.\ref{fig:syk} (b))
\begin{align}
	\mathcal{D} &=  -4.98242 \times 10^{-6} +\frac{1.13837}{V^2}, ~~~~~ \, \mathrm{for}~~~ f=1/2, \nonumber \\
	\mathcal{D} &=  -1.07932 \times 10^{-5} +\frac{0.71396}{V^2}, ~~~~~ \,\mathrm{for}~~~ f=1/5.
\end{align}
This justifies the agreement between numerical and analytical results in Fig.\ref{fig:syk} (b).

\end{document}